\documentclass[12pt]{extarticle}
\usepackage{setspace}
\usepackage{imakeidx}
\makeindex[columns=3, title=, intoc]
 
\usepackage[margin=1in]{geometry}
\usepackage{amsmath,amsthm,amssymb}
\usepackage{multicol}

\usepackage{cancel}
\usepackage{soul}
\usepackage{comment} 
\usepackage[utf8]{inputenc}
\usepackage[english]{babel}
\usepackage{textcomp}
\usepackage{amsmath}
\usepackage{gensymb}
\usepackage{multirow}
\usepackage{multicol}
\usepackage{amsbsy}
\usepackage{amssymb}
\usepackage[framemethod=TikZ]{mdframed}
\usepackage{capt-of}
\usepackage[T1]{fontenc} 
\usepackage{physics}
\usepackage{multicol} 
\usepackage{multirow}
\usepackage{lscape}
\usepackage[hang, small,labelfont=bf,up,textfont=it,up]{caption} 
\usepackage{booktabs} 
\usepackage{float} 

\usepackage{url}
\usepackage{xcolor}   
\definecolor{denim}{rgb}{0.08, 0.38, 0.74}
\definecolor{darkgreen}{rgb}{0.4, 0.8, 0.0}
\usepackage{hyperref}

\hypersetup{
    colorlinks=true,
    linkbordercolor={white},
    linkcolor={denim},
    citecolor={denim},
    filecolor={denim},      
    urlcolor={denim},
}

\usepackage{graphicx} 
\usepackage{epstopdf} 
\usepackage{paralist} 
\usepackage{subcaption}

\newcommand{\red}[1]{\textcolor{red}{#1}}
\newcommand{\blue}[1]{\textcolor{blue}{#1}}

\newcommand{\R}{\mathbb{R}}

\newcommand{\Mp}{M_{\text{Pl}}}

\newcommand{\imtau}{\text{Im}\tau}
\newcommand{\V}{\mathcal{V}}
\allowdisplaybreaks

\renewcommand{\vev}[1]{\langle #1 \rangle}

\newcommand{\action}[2]{S_{\text{#1}}^{\text{#2}}}

\newcommand{\ben}{\begin{enumerate}}
\newcommand{\een}{\end{enumerate}}

\newcommand{\bea}{\begin{eqnarray}}
\newcommand{\eea}{\end{eqnarray}}
\newcommand{\be}{\begin{equation}}
\def\bel#1{\begin{equation} \label{#1}}
\newcommand{\ee}{\end{equation}}
\newcommand{\bi}{\begin{itemize}}
\newcommand{\ei}{\end{itemize}}
\newcommand{\ba}{\begin{align}}
\newcommand{\ea}{\end{align}}

\newcommand*{\id}{{\normalfont\hbox{1\kern-0.15em \vrule width .8pt depth-.5pt}}}

\renewcommand{\footnoterule}{\vfill\kern -3pt \hrule width 0.4\columnwidth \kern 2.6pt}

\mdfdefinestyle{Question}{%
    linecolor=black!80!white,
    outerlinewidth=0.5pt,
    roundcorner=0pt,
    innertopmargin=15pt,
    innerbottommargin=15pt,
    innerrightmargin=20pt,
    innerleftmargin=20pt,
    backgroundcolor=red!10!white}
\def\question{\begin{mdframed}[style=Question]}
\def\endquestion{\end{mdframed}}

\mdfdefinestyle{Calculation}{%
    linecolor=black!80!white,
    outerlinewidth=0.5pt,
    roundcorner=0pt,
    innertopmargin=15pt,
    innerbottommargin=15pt,
    innerrightmargin=20pt,
    innerleftmargin=20pt,
    backgroundcolor=yellow!10!white}
\def\calculation{\begin{mdframed}[style=Calculation]}
    
\mdfdefinestyle{Definition}{%
    linecolor=black!80!white,
    outerlinewidth=1pt,
    roundcorner=3pt,
    innertopmargin=15pt,
    innerbottommargin=15pt,
    innerrightmargin=20pt,
    innerleftmargin=20pt,
    backgroundcolor=black!5!white}
\def\define{\begin{mdframed}[style=Definition]}

\mdfdefinestyle{ImportantEquation}{%
    linecolor=black!80!white,
    outerlinewidth=1pt,
    roundcorner=3pt,
    innertopmargin=8pt,
    innerbottommargin=12pt,
    innerrightmargin=20pt,
    innerleftmargin=20pt,
    backgroundcolor=blue!5!white}
\def\impequation{\begin{mdframed}[style=ImportantEquation]}

\def\closeframe{\end{mdframed}}

\begin{document}

\begin{titlepage}
\vskip 1 cm

\begin{center}
{\LARGE\bf \phantom{space}\par}
{\LARGE\bf \phantom{space}\par}
{\LARGE\bf A guide to frames, $2\pi$'s, scales and corrections \\ 
in string compactifications \par}
%
\vskip 1.5cm  
{\large 
%
{{B. V. Bento}}$^{a}$,   {{D. Chakraborty}}$^{b}$, {{S. Parameswaran}}$^{a}$, {{I. Zavala}}$^{c}$

\let\thefootnote\relax\footnotetext{ \hspace{-0.5cm} E-mail: {$\mathtt{Bruno.Bento@liv.ac.uk, dibya.chakraborty@ashoka.edu.in, susha@liv.ac.uk,} $}}
\let\thefootnote\relax\footnotetext{ \hspace{0.8cm}  {$\mathtt{e.i.zavalacarrasco@swansea.ac.uk} $}}
}

\vskip 0.9 cm

{\textsl{
$^{a}$ Department of Mathematical Sciences, \\ University of Liverpool, Liverpool L69 7ZL \\}
}

\vskip 0.6 cm

{\textsl{
$^{b}$ Department of Physics, Ashoka University, \\
Plot 2, Rajiv Gandhi Education City, P.O. Rai, Sonipat 131029, Haryana, India}
}

\vskip 0.6 cm

{\textsl{$^{c}$ Department of Physics, 
Swansea University,\\ Singleton Park, 
Swansea, SA2 8PP, UK}
}

\end{center}

\vskip 0.6cm

\begin{abstract}
\noindent    

\noindent This note is intended to serve as a reference for conventions used in the literature on string compactifications, and how to move between them, collected in a single and easy-to-find place, using type IIB as an illustrative example.  It will be useful to beginners in the field and busy experts.  E.g.~string constructions proposed to address the moduli stabilisation problem are generically in regions of parameter space at the boundaries of control, so that consistent use of $2\pi$'s and frame conventions can be pivotal when computing their potentially dangerous corrections.

\end{abstract}

\end{titlepage}
\pagestyle{plain}
\setcounter{page}{1}
\newcounter{bean}
\baselineskip18pt

\section{Introduction}

The idea that this pedagogical note could be a useful contribution to the community came about after several discussions with colleagues on the robustness of various candidate string constructions for moduli stabilisation, towards particle physics and cosmology, against potentially dangerous corrections.  
To scrutinise these constructions, it becomes necessary to use other people's conventions (or indeed one's own in one's past worldline), and although there is nothing deep in changing conventions, and a change of frames is simply a field redefinition\footnote{In a setup in which all couplings, including the gravitational coupling, are constant (e.g. assuming that the dilaton and volume modulus are stabilised and integrated out), the change of frames becomes simply a change of units.}$^,$\footnote{See e.g. \cite{Faraoni:1999hp, Corda:2010ye, Kamenshchik:2014waa, Karamitsos:2018lur, Bamber:2022eoy} and references therein for some interesting discussions on the equivalence of the Einstein and Jordan frames in cosmology.}, it is tedious, and possibly tricky unless starting from scratch. 
We thus present the various choices most commonly used for 10d and 4d string and Einstein frames in type IIB compactifications, and 2$\pi$'s, together with the map between them.  This should help both beginners in the field and busy experts to save some time.   We emphasise throughout the text how physical quantities such as mass ratios, which determine the size of leading corrections to explicit string compactifications, are of course convention-independent:  a change in frame would come with field redefinitions, and new masses and couplings, but the mass ratios must be invariant regardless of the choice of conventions, if a uniform choice of frames and units is used.

In Section 2, we present the 10d type IIB supergravity action in the string frame and the Einstein frame, using the two main choices of conventions for change of frames, and including the $SL(2,\mathbb{R})$ manifest action.  In Section 3, we dimensionally reduce to 4d using a general warped compactification, and present the 4d Einstein frame in different conventions which, however, always allow to recover the unwarped limit from the warped case in an intuitive way.  We identify the relations between the 4d Planck scale, (warped) string scales and (warped) KK scales.  In Section 4 we similarly work out the flux superpotential and gravitino mass, and give the mass ratio $\frac{m_{3/2}}{m_{KK}}$, which not only determines whether we have a consistent supergravity description in 4d, but also controls the higher F-term corrections to KKLT/LVS type compactifications.  In Section 5 we show the leading perturbative corrections to the K\"ahler potential and non-perturbative corrections to the superpotential in the most common conventions.   For all the relations that we present, we provide both a single expression that covers the various conventions considered commonly in the literature, and a manifestly convention-independent expression with a uniform choice of frames and units.

\section{Type IIB supergravity}

Our starting point is the type IIB low-energy supergravity action in \red{string frame}, which is given by\footnote{The action can be found on p.90 of \cite{Polchinski:1998rr}, p.314 of \cite{Becker:2007zj}, p.114 of \cite{Ibanez:2012zz}, p.79 of \cite{Baumann:2014nda} and p.625 of \cite{Blumenhagen:2013fgp}, along with the necessary definitions. There is a different definition of $F_5$, for example in \cite{Polchinski:2000uf}, which also contains the equations of motion. See also \cite{Tseytlin:1996hs} for the dilaton dependence of the RR sector.}

\begin{align}
    S_{\rm IIB}^{\red{S}} =&~ \frac{1}{2\kappa_{10}^2}\int d^{10}x\sqrt{-G^{\red{S}}}\Big\{
    e^{-2\Phi}\left(R + 4\partial_M\Phi\partial^M\Phi - \frac{1}{2}|H_3|^2\right) 
    - \left(\frac{1}{2}|F_1|^2 + \frac{1}{2}|\Tilde{F}_3|^2 + \frac{1}{4}|\Tilde{F}_5|^2\right) \Big\} \nonumber \\ 
    &- \frac{1}{4\kappa_{10}^2}\int C_4\wedge H_3 \wedge F_3 
    \label{eq:TypeIIB_action}\,,
\end{align}
where $R$ is the Ricci scalar, $\Phi$ is the dilaton, $H_3$ is the field-strength of the NS 2-form $B_2$ and $F_p$ is the field-strength of the RR $(p-1)$-forms $C_{p-1}$, and $\tilde F_3, \tilde F_5$ are defined as below:
\begin{align*}
    & H_3 = dB_2  \,,
    && \Tilde{F}_3 = F_3 - C_0 H_3 \,,\\
    & F_p = dC_{p-1} \,,
    && \Tilde{F}_5 = F_5 - \frac{1}{2}C_2\wedge H_3 + \frac{1}{2}B_2\wedge F_3 \,, \\
    & |F_p|^2 = \frac{1}{p!}F_{M_1...M_p}F^{M_1...M_p}\,.
\end{align*}
Moreover, the type IIB action must be supplemented with the self-duality condition\footnote{Notice the factor of $\frac{1}{4}$ rather than $\frac{1}{2}$ in the kinetic term, which accounts for the fact that only half the degrees of freedom should be present.}
$$\Tilde{F}_5=\star \Tilde{F}_5 \,.$$ 

The relation between the string scale $\alpha'$ and the 10d gravitational coupling in \red{string frame} $\kappa_{10}$ is 
\begin{align}
    \boxed{2\kappa_{10}^2 = (2\pi)^7\alpha'^4} \,.
\end{align}
A common convention for the string length $l_s$, which we use below, is
\be
    \boxed{(2\pi)^2\alpha' = l_s^2} \,,
\ee
although sometimes $\alpha' = l_s^2$ is used instead. The string scale (corresponding to the mass of the tower of string states) is $M_s = \frac{1}{\sqrt{\alpha'}} = \frac{2\pi}{l_s}$ for this choice of conventions. Sometimes the notation 
\begin{equation}\label{ft:stringscale}
    m_s = \frac{1}{l_s}   
\end{equation}
is used with this convention, with the relation $M_s = 2\pi m_s$.

Note that these quantities are defined in the \red{string frame}, so one should bare in mind that 
\begin{align}
    \boxed{m_s \equiv m_s^{\red{S}} 
    \quad \text{and}
    \quad l_s\equiv l_s^{\red{S}}} \,.
\end{align}
We will omit the \red{string frame} label, as commonly done in the literature, unless it is helpful to make it explicit.

\subsection{Einstein frame}
\label{sec:frames}

In the \red{string frame}, the gravitational part of the action is not in the canonical Einstein-Hilbert form. In order to obtain the latter, we perform a conformal transformation of the metric in 10d, $G^{\red{S}}\to G^{\blue{E}}=e^{2\Upsilon}G^{\red{S}}$. 

The frame in which the gravitational part of the action takes the canonical Einstein-Hilbert form -- i.e. the Ricci scalar does not couple to anything other than $\sqrt{-G^{\blue{E}}}$ -- is the \blue{Einstein frame}. This choice fixes the required conformal transformation up to a constant\footnote{Note that a constant multiplying $R^{\blue{E}}$ is a simple rescaling of the coupling constant $\kappa$, so that one still obtains the Einstein frame. The constant is a matter of convention.}
\begin{align}\label{eq:omega}
  \Upsilon = -\frac{\Phi - \Phi_0}{4}\,,
\end{align}
where the constant $\Phi_0$ is a choice of convention, and the two metrics are related by
\begin{equation}
    G^{\blue{E}}_{MN} = e^{-\frac{\Phi-\Phi_0}{2}} G^{\red{S}}_{MN} \,.
    \label{eq:changeframes}
\end{equation}
The  first term in the action \eqref{eq:TypeIIB_action}, namely the Ricci scalar, in the \blue{Einstein frame} becomes
\begin{align}
    S_{\rm grav}^{\blue{E}} 
    =& \frac{1}{2\kappa^2}\int d^{10}x~ \sqrt{-G^{\blue{E}}} 
    \Big\{ R^{\blue{E}}
    -\frac{9}{2}(G^{\blue{E}})^{MN}(\partial_M\Phi)(\partial_N\Phi) \Big\}\,,
\end{align}
where $\kappa \equiv e^{\Phi_0}\kappa_{10}$ is the rescaled coupling. Including the contribution from  the kinetic term of $\Phi$, which also transforms under this conformal transformation, 
the \blue{Einstein frame}  gravitational plus dilaton action becomes
\begin{align}
    S_{\rm grav+\Phi}^{\blue{E}} = \frac{1}{2\kappa^2}\int d^{10}x\sqrt{-G^{\blue{E}}}~ \Big\{R^{\blue{E}} - \frac{1}{2}(\partial_M\Phi)(\partial^M\Phi)\Big\} \,.
\end{align}
Note that the dilaton  is canonically normalised in \blue{Einstein frame}.

The kinetic terms of the NS and RR form fields include an implicit metric associated to the index contraction of the forms. For a generic $p$-form $\eta$ we have 
\begin{align}\label{pformE}
    |\eta|_{\red{S}}^2 
    &= \frac{1}{p!}(G^{\red{S}})^{M_1N_1}...(G^{\red{S}})^{M_pN_p}\eta_{M_1...M_p}\eta_{N_1...N_p} \nonumber \\
    &=\frac{1}{p!} (e^{2\Upsilon})^p (G^{\blue{E}})^{M_1N_1}...(G^{\blue{E}})^{M_pN_p}\eta_{M_1...M_p}\eta_{N_1...N_p} 
    = e^{2\Upsilon\cdot p}~|\eta|_{\blue{E}}^{2} \,.
\end{align}
Putting everything together, with the appropriate choice (\ref{eq:omega}), the action \eqref{eq:TypeIIB_action} in \blue{Einstein frame} becomes
\begin{align}
    S_{\rm IIB}^{\blue{E}} &= \frac{1}{2\kappa^2}\Bigg\{\int d^{10}x\sqrt{-G^{\blue{E}}}
    \left(R^{\blue{E}} - \frac{1}{2}(\partial_M\Phi)(\partial^M\Phi) - \frac{e^{\Phi_0}}{2}e^{-\Phi}|H_3|_{\blue{E}}^2\right)  \label{eq:TypeIIB} \\ 
    &-\int d^{10}x\sqrt{-G^{\blue{E}}} \left(\frac{e^{2\Phi}}{2}|F_1|_{\blue{E}}^2 
    + \frac{e^{\Phi_0}}{2}e^{\Phi}|\Tilde{F}_3|_{\blue{E}}^2 
    + \frac{e^{2\Phi_0}}{4}|\Tilde{F}_5|_{\blue{E}}^2\right)
    - \frac{e^{2\Phi_0}}{2}\int C_4\wedge H_3\wedge F_3\Bigg\}  \,. \nonumber
\end{align}
Note that the Chern-Simons term in the action does not transform, apart from via the constant relating $\kappa$ and $\kappa_{10}$, as it is a topological term, independent of the metric. The gravitational coupling is related to the string scale as
\begin{align}
    &&
    \boxed{2\kappa^2 = 2e^{2\Phi_0} \kappa_{10}^2 = (2\pi)^7 e^{2\Phi_0} \alpha'^4}
    &&
    \text{or} 
    &&
    \boxed{2\kappa^2 = \frac{e^{2\Phi_0}~l_s^8}{2\pi}} \,.
    && 
    \label{eq:string-grav-scale-convention}
\end{align}

A common  choice of $\Phi_0$ is such that the metric in the \red{string frame} and the metric in the \blue{Einstein frame} are the same at the vacuum, i.e. $\Phi_0 = \langle\Phi\rangle$ -- this allows us to discuss quantities in a frame-independent way \textit{at the vacuum}. For that choice the action in \blue{Einstein frame} reads 
\begin{align}\label{actionE}
    S_{\rm IIB}^{\blue{E}} 
    =& \frac{1}{2\kappa^2}\Bigg\{\int d^{10}x\sqrt{-G}
    \left(R - \frac{1}{2}(\partial_M\Phi)(\partial^M\Phi) - \frac{g_s}{2}e^{-\Phi} |H_3|^2\right) \nonumber \\ 
    &-\int d^{10}x\sqrt{-G} \left(\frac{e^{2\Phi}}{2}|F_1|^2 + \frac{g_s}{2}e^{\Phi}|\Tilde{F}_3|^2 + \frac{g_s^2}{4}|\Tilde{F}_5|^2\right)
    - \frac{g_s^2}{2}\int C_4\wedge H_3\wedge F_3\Bigg\},
\end{align}
where we dropped the \blue{E}, as all metrics are in \blue{Einstein frame}. 
With this choice, the gravitational coupling is related to the string scale as 
\begin{align}
    &&
    2\kappa^2 = 2g_s^2\kappa_{10}^2 = (2\pi)^7g_s^2\alpha'^4
    &&
    \text{or} 
    &&
    2\kappa^2 = \frac{g_s^2~l_s^8}{2\pi} \,.
    && 
\end{align}

Another common choice of convention is $\Phi_0=0$.  In this case, volumes are frame-dependent in the vacuum (see eq. \eqref{eq:Vd_between_frames} below) and one needs to be careful to compare quantities consistently, e.g. when checking whether the $\alpha'$-expansion (usually computed in \red{string frame}) is under control for a certain vacuum (usually obtained in \blue{Einstein frame}). 
For this choice the gravitational coupling is related to the string scale as 
\begin{align}
  &&
  2\kappa^2 = 2\kappa_{10}^2 = (2\pi)^7\alpha'^4
  &&
  \text{or} 
  &&
  2\kappa^2 = \frac{l_s^8}{2\pi} \,.
  &&
\end{align}

It might seem strange that this relation between scales (the gravitational scale and the string scale) is convention-dependent; as we will emphasise throughout this note, mass ratios should \textit{not} depend on convention. The convention dependence in \eqref{eq:string-grav-scale-convention} arises because we are comparing scales measured in different frames---while $\kappa$ is the \blue{Einstein frame} gravitational coupling, $\alpha'$ is the \red{string frame} string scale. The mass of any state measured in the \blue{Einstein frame} satisfies 
\begin{equation}
    (m^{\blue{E}})^2 = (G^{\blue{E}})^{MN}p_M p_N = e^{\frac{\Phi-\Phi_0}{2}} (G^{\red{S}})^{MN} p_M p_N = e^{\frac{\Phi-\Phi_0}{2}} (m^{\red{S}})^2 \,.
\end{equation}
This is in particular true for the string mass $M_s = 1/ \sqrt{\alpha'}=2\pi/\ell_s$, or in terms of $\alpha'$ and $l_s$, 
\begin{align}
    \boxed{\alpha'_{\blue{E}} = e^{\frac{\Phi-\Phi_0}{2}} \alpha'_{\red{S}}  
    \quad\text{and} 
    \quad l_s^{\blue{E}} = e^{-\frac{\Phi-\Phi_0}{4}}l_s^{\red{S}}} \,.
    \label{eq:string-scale-einstein-string}
\end{align}
Therefore, when we compare \blue{Einstein frame} quantities we find the convention-independent relations
\begin{align}
    &&
    \boxed{2\kappa^2 = (2\pi)^7 e^{2\Phi} (\alpha'_{\blue{E}})^4}
    &&
    \text{or} 
    &&
    \boxed{2\kappa^2 = \frac{e^{2\Phi}~(l_s^{\blue{E}})^8}{2\pi}} \,.
    && 
    \label{eq:string-grav-scale-convention-independent}
\end{align}

It is worth emphasising that volumes measured using a \red{string frame} metric may differ from the ones measured using an \blue{Einstein frame} metric, depending on the convention used, i.e. on the choice of $\Phi_0$. 
Using the relation between the two metrics \eqref{eq:changeframes}, a generic $d$-dimensional volume can be written as
\begin{equation}\label{eq:volume-frames}
    V_d^{\blue{E}} = \int d^d y \sqrt{g_d^{\blue{E}}} ~f(y)
    = \int d^d y \sqrt{g_d^{\red{S}}} ~e^{-\frac{d}{4}(\Phi - \Phi_0)} f(y) \,,
\end{equation}
where we allow for some function $f(y)$, such as $H(y)$ in the definition of the warped volume $V_{\rm w}$, (\ref{eq:Vw}) below. It follows that the volumes in the two frames (assuming $\Phi$ is stabilised) are related as 
\begin{equation}
    V_d^{\red{S}} = \big(e^{\langle\Phi\rangle-\Phi_0}\big)^{\frac{d}{4}} V_d^{\blue{E}}. 
\end{equation}
In \red{string frame} string length units, $l_s^{\red{S}}$, and for the particular case with $d=6$, dimensionless volumes are related as
\begin{equation}\label{eq:Vd_between_frames}
  \boxed{  \V_{\red{S}} = e^{\frac{3}{2}(\langle\Phi\rangle-\Phi_0)}\V_{\blue{E}}} \,.
\end{equation}
Hence, it is important to note the convention being used for the \blue{Einstein frame} metric and be consistent when comparing quantities  that are obtained in either \red{string frame} or \blue{Einstein frame} (e.g. perturbative and non-perturbative corrections). 
It will also be helpful to define another dimensionless \blue{Einstein frame} volume, $\hat{\mathcal{V}}_{\blue{E}}$, which is given in \blue{Einstein frame} string length units \eqref{eq:string-scale-einstein-string} and is related to $\mathcal{V}_{\red{S}}$ as
\begin{equation}\label{eq:Einstein-einstein-volume}
    \boxed{\hat{\mathcal{V}}_{\blue{E}} = \mathcal{V}_{\red{S}}} \,.  
\end{equation}

\subsection{\texorpdfstring{$SL(2,\R)$}{TEXT} manifest action}
\label{sec:SL(2,R)action}

We now express the   \blue{Einstein frame} action \eqref{actionE}  in terms of the fields $G_3$ and $\tau$, such that the underlying $SL(2,\R)$ symmetry becomes manifest, which is sometimes useful when doing calculations and it is commonly used in the literature. We define the fields 
\begin{align}
   \tau &= C_0 + ie^{-\Phi}\,, \\ 
   G_3 &= \Tilde{F}_3 - ie^{-\Phi}H_3 = F_3 - \tau H_3\,,
\end{align}
where $\tau$ is known as the \textit{axio-dilaton}.

In terms of these fields,  the action takes the form 
\begin{align}
    S_{\rm IIB}^{\blue{E}} 
    =& ~ \frac{1}{2\kappa^2}\int d^{10}x\sqrt{-G}
    \left(R 
    - \frac{(\partial_M\tau)(\partial^M\Bar{\tau})}{2(\Im\tau)^2} 
    - \frac{e^{\Phi_0}}{2(\Im\tau)}|G_3|^2 
    - \frac{e^{2\Phi_0}}{4}|\Tilde{F}_5|^2\right) \nonumber \\ 
    &- \frac{1}{2\kappa^2}\frac{ie^{2\Phi_0}}{4}\int \frac{1}{(\Im\tau)} C_4\wedge G_3\wedge \overline{G}_3 \,,
    \label{eq:TypeIIB_SL2Z}
\end{align}
where we recall $\kappa = e^{\Phi_0}\kappa_{10}$. We can also write the action in differential form language,
\begin{align}
    S_{\rm IIB}^{\blue{E}} 
    =&  \frac{1}{2\kappa^2}\int 
    \Big(R\star 1 - \frac{d\tau\wedge\star d\tau}{2(\Im\tau)^2} 
    - \frac{e^{\Phi_0}}{2(\Im\tau)}G_3\wedge\star\overline{G}_3 
    - \frac{e^{2\Phi_0}}{4}\Tilde{F}_5\wedge\star\Tilde{F}_5 \nonumber \\
    &- \frac{ie^{2\Phi_0}}{4(\Im\tau)}C_4\wedge G_3\wedge \overline{G}_3\Big).
\end{align}
Written in this form, the $SL(2,\R)$ symmetry of the type IIB action becomes manifest --- it leaves the metric and $4$-form invariant, and acts on the remaining fields as 
\begin{align}
    \tau\to \frac{a\tau + b}{c\tau + d} \,, 
    && \begin{pmatrix}
     C_2 \\
     B_2
    \end{pmatrix} = 
    \begin{pmatrix}
     a & b \\
     c & d
    \end{pmatrix}\begin{pmatrix}
      C_2 \\ 
      B_2
    \end{pmatrix} \,,
    && \text{with}\quad 
    \begin{pmatrix}
     a & b \\
     c & d
    \end{pmatrix}\in SL(2,\R) \,,
\end{align}
that is, $ad-bc=1$. 

Another occasionally used convention (see e.g. \cite{Tonioni:2022izq,McGuirk:2012sb,Martucci:2005rb,Myers:1999ps}) is to redefine the RR forms in \blue{Einstein frame} as $C_p^{\blue{E}} = e^{\Phi_0} C_p^{\red{S}}$; the action then becomes
\begin{align}
    S_{\rm IIB}^{\blue{E}} 
    = \frac{1}{2\kappa^2}\int 
    \Big(R\star 1 - \frac{d\tau\wedge\star d\tau}{2(\Im\tau)^2} 
    - \frac{G_3\wedge\star\overline{G}_3}{2(\Im\tau)} 
    - \frac{1}{4}\Tilde{F}_5\wedge\star\Tilde{F}_5 
    - \frac{i}{4(\Im\tau)}C_4\wedge G_3\wedge \overline{G}_3\Big) \,,
\end{align}
where the axio-dilaton was also redefined as $\tau^{\blue{E}} = e^{\Phi_0} \tau^{\red{S}} = e^{\Phi_0} C_0^{\red{S}} + i e^{-\varphi}$, with $e^{-\varphi} = e^{-(\Phi-\Phi_0)}$, and $G_3^{\blue{E}} = e^{\Phi_0} G_3^{\red{S}} = F_3^{\blue{E}} - \tau^{\blue{E}} H_3$. Note that in terms of $\tau^{\blue{E}}$ we have $\langle\Im\tau^{\blue{E}}\rangle = 1$. With this field redefinition the action looks the same regardless of the choice of convention, apart from having a different gravitational coupling $\kappa$.

\section{Dimensional Reduction}
\label{sec:scalar_potential}

In order to obtain a 4d EFT at low energies, we consider a compactification (or dimensional reduction) of the 10d theory down to 4 dimensions. The 4d theory describes perturbations around a 10d vacuum solution and is valid for energies much lower than the compactification scale\footnote{Depending on the details of the compactification, this scale could correspond to e.g. $m_{KK}$ or $m_{KK}^{\rm w}$.}. We consider a vacuum solution which corresponds to a warped product spacetime $\mathcal{M}_{10} = \R^{1,3}\times_{\rm w} X_6$, where $\R^{1,3}$ is a 4d Lorentzian spacetime and $X_6$ is a 6d compact space. 

The \blue{10d Einstein frame} metric is given by
\begin{equation}
    ds_{10}^2 = H^{-1/2}(y)~e^{2\omega(x)}g_{\mu\nu} dx^\mu dx^\nu
    + H^{1/2}(y)~\V^{1/3} g_{mn} dy^m dy^n \,,
    \label{eq:10dmetric}
\end{equation}
where $x^\mu ~(\mu=0,...,3)$ are 4d coordinates and $y^m ~(m=4,...,9)$ are 6d coordinates on the compact space $X_6$. 
The metric $g_{mn} = (g_6)_{mn}$ is the 6d metric of a Calabi-Yau (Ricci flat) manifold usually normalised as
$$\int d^6 y\sqrt{g_6} \equiv (l_s^{\red{S}})^6 \,.$$
so that $\V$ keeps track of the physical size of the compact space in the \blue{Einstein frame} using \red{string frame} string length units and thus we can also call it (see \eqref{eq:Vd_between_frames})
\begin{equation}
\V_{\blue{E}}\equiv\V\,.
\end{equation}
We define the warp factor $H$ as 
\be\label{eq:warpfactor}
    H(y) \equiv 1 + \frac{e^{-4A_0(y)}}{\V^{2/3}} \,,
\ee
\noindent which is motivated as follows. First, the background warp factor -- commonly written as $e^{-4A(y)}$ -- that solves the 10d Einstein equations in the presence of fluxes is only fixed up to a constant shift, $e^{-4A(y)} = e^{-4A_0(y)} + c$, which becomes a modulus in the 4d EFT \cite{Frey:2008xw}.
The fact that $g_{mn}\rightarrow\lambda g_{mn}$ together with $e^{2A}\rightarrow \lambda e^{2A}$ is a gauge redundancy of the metric \cite{Giddings:2001yu,Giddings:2005ff,Aparicio:2015psl} allows us to choose $\lambda=c^{1/2}$ and rewrite $e^{-4A(y)} = 1 + \frac{e^{-4A_0(y)}}{c}$, which naturally recovers the unwarped case in the $c\rightarrow\infty$ limit -- this relates $c=\mathcal{V}^{2/3}$ with the unwarped volume of the compact space. The factor $e^{2\omega(x)}$ is introduced to Weyl rescale to the \blue{4d Einstein frame}, with metric $g_{\mu\nu}=g_{\mu\nu}^{\blue{E_4}}$, as we now describe.

Dimensionally reducing the 10d Einstein-Hilbert term (in \blue{10d Einstein frame})
\begin{equation}
    S_{\rm IIB}^{\blue{E}} 
    = \frac{1}{2\kappa^2}\int d^{10}x\sqrt{-G} 
    ~R_{10} 
    \label{eq:10EH}
\end{equation}
down to 4d using the ansatz (\ref{eq:10dmetric}) gives, among other contributions, the term 
\begin{align}\label{eq:JordanFrame}
    S_{4d} 
    &\supset \frac{1}{2\kappa^2}
    \int d^{4}x \sqrt{-g_4} \cdot  e^{2\omega(x)}\Big( \V
    \int d^{6}y \sqrt{g_6} \cdot H(y) \Big) R_4   \,. 
\end{align}
Any choice of $e^{2\omega(x)}$ that leaves a non-canonical coupling of the volume modulus $\V$ to $R_4$ is said to be in the \red{Jordan frame}. Requiring a canonical form for the Einstein-Hilbert term instead -- which defines the \blue{4d Einstein frame} -- fixes the Weyl rescaling $e^{2\omega(x)}$, up to a constant factor $e^{2\omega_0}$, as 
\begin{equation}
    \label{eq:4d-weyl-rescaling}
    e^{2\omega(x)} = \frac{e^{2\omega_0}\cdot l_s^6}{\V
    \int d^{6}y \sqrt{g_6} \cdot H(y)} \equiv \frac{e^{2\omega_0}\cdot l_s^6}{V_{\rm w}} = \frac{e^{2\omega_0}}{\V_{\rm w}}\,,
\end{equation}
where we defined the \blue{Einstein frame} \textit{warped volume} $V_{\rm w} = \V_{\rm w}\cdot l_s^6$, in \red{string frame} string length units, as
\begin{equation}
    V_{\rm w} \equiv \V
    \underbrace{\int d^{6}y \sqrt{g_6} \cdot H(y)}_{\langle H \rangle_{\text{av}}\cdot ~l_s^6} \,.
    \label{eq:Vw}
\end{equation}
This definition of $V_{\rm w}$ only differs from $\V\cdot l_s^6$ by the factor $\langle H \rangle_{\text{av}}$, the average of the warp factor over the compact space. If the integral is dominated by the unwarped bulk, then $\langle H \rangle_{\text{av}}\approx 1$ and $V_{\rm w} \approx \V\cdot l_s^6$.  

Note the similarities with the conformal transformation in 10d to go from \red{string frame} to \blue{Einstein frame}, where we also had some freedom in the form of a constant. We saw there that a convenient choice was the one for which the two metrics matched \textit{at the vacuum}. Here we are going from the \red{Jordan frame}, in which some scalars couple to the Ricci scalar in the action, to the \blue{4d Einstein frame}, in which we recover the canonical Einstein-Hilbert term. The two metrics will match \textit{at the vacuum} if we choose $e^{2\omega_0}= \langle \V_{\rm w} \rangle$. The action in the \blue{4d Einstein frame} for general $\omega_0$ becomes
\begin{align}
    S_{4d}^{\blue{E}} 
    &\supset \frac{e^{2\omega_0}\cdot l_s^6}{2\kappa^2}
    \int d^{4}x \sqrt{-g_4} \cdot  R_4 
    \equiv \frac{M_\text{Pl}^2}{2} \int d^{4}x \sqrt{-g_4} \cdot  R_4  \,, 
\end{align}
which defines the relation between the string scale $(m_s = 1/l_s)$ and the Planck scale as
\begin{equation}\label{eq:msMp-relation}
    \boxed{m_s = \frac{e^{\Phi_0}}{\sqrt{4\pi e^{2\omega_0}}} \Mp} \,.
\end{equation}
For the convenient choice $e^{\Phi_0} = g_s$ and $e^{2\omega_0} = \langle \V_{\rm w} \rangle$, this relation becomes
\begin{equation}
    m_s = \frac{g_s}{\sqrt{4\pi \V_{\rm w}}} \Mp \,.
\end{equation}
Note also that in the unwarped limit the warped volume tends to the volume modulus of the compactification, $\V_{\rm w}\to \V$, and -- with these choices of convention for the Weyl rescalings -- we recover the common expression for the ratio $m_s/\Mp$.  If instead we choose conventions $\Phi_0=0=\omega_0$, then
\begin{equation}
    m_s = \frac{\Mp}{\sqrt{4\pi}} \,.
\end{equation}

This convention dependence arises again from the fact that we are comparing quantities measured in different frames, as $m_s$ is the string mass measured in the \red{string frame}. 
Analogously to the 10d change of frames, it is now useful to introduce the \blue{4d Einstein frame} string scale in order to obtain a manifestly convention-independent relation.  Consider first string states localised in regions of constant warp factor, $H(y_0)$ (these states could arise, for example, from open string states localised on a brane at the tip of a warped throat).  The \blue{4d Einstein frame} mass for such states is given by
\begin{align}
    \left(\left(m_{s}^{{\rm w}}\right)^{\blue{E_4}}\right)^2 
    &= g^{\mu\nu}_{\blue{E_4}} p_\mu p_\nu 
    = H(y_0)^{-1/2}\cdot\frac{e^{2\omega_0}}{\V\cdot \langle H \rangle_{\text{av}}}\cdot G^{\mu\nu}_{\blue{E}} p_\mu p_\nu \nonumber \\
    &= \frac{H(y_0)^{-1/2}}{ \langle H \rangle_{\text{av}}}\cdot\frac{e^{2\omega_0}}{\hat{\V}_{\blue{E}}~e^{-\frac{3}{2}(\langle\Phi\rangle-\Phi_0)}}\cdot e^{\frac{\langle\Phi\rangle-\Phi_0}{2}}\cdot G^{\mu\nu}_{\red{S}} p_\mu p_\nu \nonumber \\
    &=  \frac{H(y_0)^{-1/2}}{ \langle H \rangle_{\text{av}}} \cdot \frac{e^{2\langle\Phi\rangle}}{\hat{\V}_{\blue{E}}}\cdot\frac{e^{2\omega_0}}{e^{2\Phi_0}} \cdot (m_s^{\red{S}})^2 \nonumber \\
    &= \frac{H(y_0)^{-1/2}}{ \langle H \rangle_{\text{av}}}\cdot \frac{e^{2\langle\Phi\rangle}}{4\pi\hat{\V}_{\blue{E}}}\Mp^2 \,.
    \label{eq:mass-4dEinstein-frame}
\end{align}
where recall that $\hat{\V}_{\blue{E}}$ is the \blue{Einstein frame} volume in \blue{Einstein frame} string length units, $\ell_s^{\blue{E}}$.  It follows from \eqref{eq:mass-4dEinstein-frame} that the \textit{warped} string scale for localised states is given by 
\be
    \boxed{(m_{s}^{\rm w})^{\blue{E_4}} \equiv \frac{H(y_0)^{-1/4}}{\langle H \rangle_{\text{av}}^{1/2}} \cdot\frac{e^{\langle\Phi\rangle}}{\sqrt{4\pi\hat{\V}_{\blue{E}}}} \Mp} \,, \label{E:mswE}
\ee 
whose relation to $\Mp$ is fully convention independent. 

For closed string states that are allowed to propagate along directions of varying warp factor, one needs to take into account their dimensional reduction.   Let us take the scalar field $\psi$ to represent a closed string state of mass $m_s^{\red{S}}$ measured in the \red{string frame} (here and below we drop overall factors that would simply be absorbed in the canonical normalisation),  
\begin{align}
    S &\propto \int d^{10}x \sqrt{-G^{\red{S}}} \Big\{
        -\frac{1}{2}G^{MN}_{\red{S}} (\partial_M \psi)(\partial_N \psi) + (m_s^{\red{S}})^2 \psi^2
    \Big\} \nonumber \\
    &\propto \int d^{10}x \sqrt{-G^{\blue{E}}} \Big\{
        -\frac{1}{2} e^{\frac{\langle\Phi\rangle - \Phi_0}{2}}G^{MN}_{\blue{E}} (\partial_M \psi)(\partial_N \psi) + (m_s^{\red{S}})^2 \psi^2
    \Big\} \nonumber \\
    &= \int d^4x ~e^{4\omega(x)} \sqrt{-g_4} \int d^6y \sqrt{g_6} ~ H^{1/2}(y) \V  \Big\{
        -\frac{1}{2} e^{\frac{\langle\Phi\rangle - \Phi_0}{2}}H^{1/2}(y) e^{-2\omega(x)}g^{\mu\nu} (\partial_\mu \psi)(\partial_\nu \psi) \nonumber \\
        &\hspace{8cm} 
        + \text{internal momentum}
        + (m_s^{\red{S}})^2 \psi^2
    \Big\} \nonumber \\ 
    &= \int d^4x \sqrt{-g_4} ~e^{\frac{\langle\Phi\rangle - \Phi_0}{2}} e^{2\omega(x)} \V \int d^6y \sqrt{g_6} ~\Big\{
        -\frac{1}{2}~H(y) g^{\mu\nu} (\partial_\mu \psi)(\partial_\nu \psi) 
        + \text{internal momentum} \nonumber \\
        &\hspace{7.5cm} 
        + e^{-\frac{\langle\Phi\rangle - \Phi_0}{2}}e^{2\omega(x)} H^{1/2}(y)(m_s^{\red{S}})^2 \psi^2
    \Big\}    \,.
\end{align}
Next, it is useful to decompose the effective 10d field into a tower of 4d KK modes, with
\begin{equation} 
    \psi(x,y) = \sum_k \psi_k(x)\xi^k(y) \,,
\end{equation}
where the eigenmodes $\xi^k$ satisfying the orthonormality condition
\begin{equation} \label{eq:orthonormal}
    \int d^6y \sqrt{g_6}\cdot H(y) ~\xi^k\xi^l =  \ell_s^6 \cdot \delta^{kl} \,.
\end{equation}
Then using \eqref{eq:4d-weyl-rescaling}, we find for the lowest mode $\psi_0$ in the tower, which has constant $\xi^0 = \frac{1}{\langle H\rangle_{\rm av}^{1/2}}$,
\begin{align}
    S &\propto \int d^4x \sqrt{-g_4}\Big\{
        -\frac{1}{2} g^{\mu\nu} (\partial_\mu \psi_0)(\partial_\nu \psi_0) 
        \cdot \int d^6y \sqrt{g_6}\cdot H(y) ~(\xi^0)^2 \nonumber \\
        &\hspace{2cm} 
        + e^{-\frac{\langle\Phi\rangle - \Phi_0}{2}}\cdot\frac{e^{2\omega_0}}{\V_{\blue{E}}\cdot\langle H \rangle_{\rm av}}\cdot(m_s^{\red{S}})^2 \psi_0^2 
        \cdot \underbrace{\int d^6y \sqrt{g_6}\cdot H(y)^{1/2}~(\xi^0)^2}_{\frac{\langle H^{1/2} \rangle_{\rm av}}{\langle H\rangle_{\rm av}} \cdot \ell_s^6}
    \Big\} \,,  \nonumber \\ 
    &\propto \int d^4x \sqrt{-g_4}\Big\{
        -\frac{1}{2} g^{\mu\nu} (\partial_\mu \psi_0)(\partial_\nu \psi_0) 
        + (m_s^{\blue{E_4}})^2 \psi_0^2 
    \Big\} \,,
\end{align}
and identify the \blue{4d Einstein frame} mass for the (canonically normalised) string state as
\begin{equation}
    (m_s^{\blue{E_4}})^2 =\frac{\langle H^{1/2} \rangle_{\rm av}}{\langle H \rangle_{\rm av}^{2}} \cdot \frac{e^{2\langle\Phi\rangle}}{\hat{\V}_{\blue{E}}}\cdot\frac{e^{2\omega_0}}{e^{2\Phi_0}} 
    \cdot(m_s^{\red{S}})^2 \,.
\end{equation}
After using \eqref{eq:msMp-relation}, we see that the relation between the \blue{4d Einstein frame} string scale, $m_s^{\blue E_4}$, and $\Mp$ is manifestly convention-independent.  For example, if we assume that most of the compact space is unwarped  -- or in the absence of warping, $H=1$ -- we find
\begin{equation}
    \boxed{m_s^{\blue{E_4}} = \frac{e^{\langle\Phi\rangle}}{\sqrt{4\pi\hat{\V}_{\blue{E}}}} \Mp} \,. \label{E:msE}
\end{equation}
Moreover, we recover the standard relation between the warped string scale for localised states \eqref{E:mswE} and the bulk string scale \eqref{E:msE} as
\begin{equation}
	\boxed{(m_{s}^{\rm w})^{\blue{E_4}} = H^{-1/4}(y_0) \cdot m_s^{\blue{E_4}}}  \,.
\end{equation}		

Finally, let us determine the Kaluza-Klein (KK) scale at which the towers of massive states associated with the compact dimensions appear. Considering the simple case of a 10d scalar field $\rho$,
\begin{align}
    S &\propto \int d^{10}x \sqrt{-G_{\blue{E}}} \left\{-\frac{1}{2}G^{MN}_{\blue{E}}(\partial_M\rho)(\partial_N\rho) \right\} \nonumber \\
    &= \int d^4x \int d^6y \cdot H^{-1}(y)e^{4\omega(x)}\sqrt{-g_4}\cdot H^{3/2}(y)~\V\sqrt{g_6} \nonumber \\
    &\hspace{3em} \left\{-\frac{1}{2}H^{1/2}(y) e^{-2\omega(x)}g^{\mu\nu}(\partial_\mu\rho)(\partial_\nu\rho) 
    -\frac{1}{2}H^{-1/2}(y)\V^{-1/3}g^{mn}(\partial_m\rho)(\partial_n\rho)\right\} \nonumber \\
    &= \int d^4x \sqrt{-g_4} ~ e^{2\omega }~ \V \int d^6y \sqrt{g_6} \left\{-\frac{1}{2}H(y)g^{\mu\nu}(\partial_\mu\rho)(\partial_\nu\rho) 
    -\frac{1}{2}\frac{e^{2\omega(x)}}{\V^{1/3}}g^{mn}(\partial_m\rho)(\partial_n\rho)\right\} \nonumber \\
    &= \int d^4x \sqrt{-g_4} ~ e^{2\omega} ~ \V \int d^6y \sqrt{g_6} \left\{-\frac{1}{2}H(y)g^{\mu\nu}(\partial_\mu\rho)(\partial_\nu\rho) 
    +\frac{1}{2}\frac{e^{2\omega(x)}}{\V^{1/3}}H(y)(\Delta_6\rho)\cdot\rho\right\} 
    \,,
\end{align}
where in the last step we integrated the second term by parts and defined the internal space Laplacian operator
\begin{equation}\label{eq:Laplacian}
    \Delta_6 \rho \equiv \frac{H^{-1}(y)}{\sqrt{g_6}}\partial_m(\sqrt{g_6}~g^{mn}\partial_n\rho) \,.
\end{equation}
Decomposing the field $\rho(x,y)$ in a basis of eigenfunctions of $\Delta_6$ (i.e. $\Delta_6\xi^k = - \lambda_k^2 \xi^k$, with no sum over $k$), 
\begin{equation}
    \rho(x,y) = \sum_k \varrho_k(x)\xi^k(y) \,,
\end{equation}
the action for the 10d scalar $\rho$ becomes an action for infinitely many 4d scalars $\varrho_k(x)$,
\begin{align}
    S &\propto \int d^4x \sqrt{-g_4} ~ \V \int d^6y \sqrt{g_6} 
    \sum_{k,l} \Bigg\{-\frac{1}{2}H(y)g^{\mu\nu}(\partial_\mu\varrho_k)(\partial_\nu\varrho_l)(\xi^k\xi^l) \nonumber \\
    &\hspace{15em} 
    -\frac{1}{2}\frac{e^{2\omega(x)}}{\V^{1/3}}H(y)\lambda_k^2 ~\varrho_k\varrho_l ~(\xi^k\xi^l)\Bigg\} \nonumber \\
    &= \int d^4x \sqrt{-g_4} ~ \V \sum_{k,l}
     \Bigg\{-\frac{1}{2}g^{\mu\nu}(\partial_\mu\varrho_k)(\partial_\nu\varrho_l)
     \cdot\int d^6y \sqrt{g_6}\cdot H(y) ~ \xi^k\xi^l \nonumber \\
    &\hspace{11em} 
    -\frac{1}{2}\frac{e^{2\omega(x)}}{\V^{1/3}}\lambda_k^2\varrho_k\varrho_l 
    \cdot\int d^6y \sqrt{g_6}\cdot H(y) ~\xi^k\xi^l \Bigg\} \,.
\end{align}
Since the eigenmodes $\xi^k$ satisfy the orthonormality condition \eqref{eq:orthonormal}, 
the KK modes decouple and the action reduces to
\begin{align}
    S &\propto \sum_{k}\int d^4x \sqrt{-g_4} ~  
     \Bigg\{-\frac{1}{2}g^{\mu\nu}(\partial_\mu\varrho_k)(\partial_\nu\varrho_k)(\V)
    +\frac{1}{2}\frac{e^{2\omega(x)}}{\V^{1/3}}\lambda_k^2\varrho_k^2 
    (\V)\Bigg\} \nonumber \\
    &\propto \sum_{k}\int d^4x \sqrt{-g_4} ~  
     \Bigg\{-\frac{1}{2}g^{\mu\nu}(\partial_\mu\varrho_k)(\partial_\nu\varrho_k)
    +\frac{1}{2}\cdot\frac{\lambda_k^2}{\V^{1/3}}\frac{e^{2\omega_0}}{\V_{\rm w}}\cdot(\varrho_k)^2 \Bigg\}
    \,.
\end{align}
Therefore, the mass of (the canonically normalised) $\varrho_k$ is 
\begin{equation} \label{eq:mk}
    m_k = \frac{\lambda_k}{\V^{1/6}}\left(\frac{e^{2\omega_0}}{\V_{\rm w}}\right)^{1/2}
    \quad\implies\quad
    m_{KK} = \frac{\lambda_1}{\V^{1/6}}\left(\frac{e^{2\omega_0}}{\V_{\rm w}}\right)^{1/2}
    \,,
\end{equation}
where we identify the \blue{4d Einstein frame} KK scale, $m_{KK}$, with the mass of the lightest mode $m_1$.   To determine $\lambda_k$ (and the eigenfunctions\footnote{These are commonly referred to as the wavefunctions of the modes $\varrho_k$.}) one must solve the eigenvalue equation
\be\label{eq:eigenvalue}
    \frac{1}{\sqrt{g_6}}\partial_m(\sqrt{g_6}~g^{mn}\partial_n \xi^k) + H(y)\cdot\lambda_k^2\cdot\xi^k = 0 \,,
\ee 
together with appropriate boundary conditions. It is therefore not possible to give a fully generic expression for $\lambda_k$, and thus $m_{KK}$, as it depends on the details of the compactification, but we will consider two illustrative examples, giving a `bulk' KK scale and a `warped' KK scale.

First, let us consider the case of a torus with a single common radius as the prototypical example of an isotropic compact space\footnote{More generically one could consider different scales in different directions, which would result in different KK scales.} with characteristic scale $\ell_s$\footnote{This is consistent with our normalisation for the coordinates $y^m$, such that $\int d^6y \sqrt{g_6} = l_s^6$ -- it corresponds to an identification of the normalised coordinates $y^m\sim y^m + 1$ (with $ds_6^2 = l_s^2~ dy^mdy_m$), rather than $y^m\sim y^m + 2\pi$. Moreover, the eigenfunctions respecting the (periodic) boundary conditions on the torus would be $\xi^{\vec{k}} \propto e^{2\pi i ~\vec{k}\cdot\vec{y}}$, with a vector of integers $\vec{k}$ labeling the modes, and hence $\Delta_6\xi^{\vec{k}} = - (2\pi)^2 k^2 \cdot m_s\cdot \xi^{\vec{k}}$, giving $\lambda_{\vec{k}} = (2\pi |\vec{k}|)\cdot m_s$.}.  We will assume moreover that the eigenfunctions have non-trivial amplitude throughout the bulk, and that most of the compact space is unwarped.   Therefore, in solving \eqref{eq:eigenvalue}, we can approximate $H(y) \approx 1$ and the eigenvalue becomes $\lambda_1 \approx (2\pi)\cdot m_s$. Using this result in \eqref{eq:mk} leads to a KK scale:
\begin{equation}\label{eq:mKK}
    \boxed{
    m_{KK} \approx \left(\frac{e^{2\omega_0}}{\V}\right)^{1/2} \frac{2\pi}{\V^{1/6}}m_s 
    = \sqrt{\frac{\pi}{2}}\cdot \frac{e^{\Phi_0}}{\sqrt{4\pi}\V^{2/3}} \Mp} \,.
\end{equation}
For the convenient choice $e^{\Phi_0} = g_s$ and $e^{2\omega_0} = \V$ we have
\begin{align}\label{eq:mKKconv}
    m_{KK} \approx
    \frac{2\pi}{\V^{1/6}}m_s
    = \sqrt{\frac{\pi}{2}} \cdot \frac{g_s}{\V^{2/3}}\Mp \,,
\end{align}
while for the common alternative choice $\Phi_0=0$, the factor of $g_s$ will be absent.  Once again, the mass relations are convention-dependent because we are mixing different frames and units.  Using the \blue{4d Einstein frame} for all the masses and \blue{Einstein frame} string length units for the \blue{Einstein frame} volumes yields manifestly convention-independent relations,
\begin{equation}\label{eq:mKKindi}
    \boxed{
    m_{KK} \approx \frac{2\pi}{\hat{\V}_{\blue{E}}^{2/3}} m_s^{{\blue{E_4}}} 
    = \sqrt{\frac{\pi}{2}}\cdot \frac{e^{\langle\Phi\rangle}}{\sqrt{4\pi}\hat{\V}_{\blue{E}}^{2/3}} \Mp} \,.
\end{equation}

Next, consider a warped compactification in which the warping dominates and the KK modes become localised in the region of maximum warping, $y\sim y_0$, where their energies are minimised\footnote{The same result is obtained for the KK reduction of $(p+1)$-dimensional fields descending from open strings that are localised on a subspace at some fixed $y=y_0$, e.g. fields living on the world-volume of a $p$-brane wrapping a $(p-3)$-internal cycle somewhere along a warped throat in 10d.}.  Now in solving \eqref{eq:eigenvalue}, we can approximate the warp factor as $H(y) \sim H(y_0)$.  Then we can use $\lambda_k \sim H(y_0)^{-1/2} \cdot m_s$ in \eqref{eq:mk}, which leads to a KK scale
\begin{equation}\label{eq:mKKloc}
    \boxed{
    m_{KK}^{\rm w} \sim \left(\frac{e^{2\omega_0}}{\V \cdot\langle H\rangle_{\rm av}}\right)^{1/2} \frac{1}{H(y_0)^{1/2}\V^{1/6}}\cdot m_s 
    = \frac{1}{\langle H\rangle_{\rm av}^{1/2}}\cdot \frac{e^{\Phi_0}}{\sqrt{4\pi}H(y_0)^{1/2}\V^{2/3}} \Mp} \,.
\end{equation}
For example, in the well-studied geometry of a Klebanov-Strassler (KS) warped throat \cite{Klebanov:2000hb}, one has at the tip $R_{S^3} \equiv \mathcal{R}_{S^3} \ell_s = H(y_0)^{1/4} \V^{1/6} \ell_s \approx e^{-A(y_0)} \ell_s$, where moreover $R_{S^3} \approx \sqrt{\alpha' g_s M}$.  Then, choosing the convention $e^{2\omega_0}=1$ and assuming $\V \cdot \langle H \rangle_{\rm av} \approx \V$, we recover the familiar result for the warped KK scale at the tip of the KS throat (see e.g. \cite{DeWolfe:2002nn, Douglas:2008jx, Aparicio:2015psl} and eq.~(2.26) in \cite{Burgess:2006mn})
\begin{equation}
    m_{KK}^{\rm w} \sim \frac{1}{\V^{1/2}H(y_0)^{1/4}R_{S^3}} = \frac{e^{A(y_0)}}{\V^{1/3}} \cdot \frac{1}{{R}_{S^3}} \,.
\end{equation}
Finally, we write down manifestly convention-independent relations by using \blue{Einstein frame} quantities throughout,
\begin{equation}\label{eq:mKKms}
    \boxed{
    m_{KK}^{\rm w} \sim \frac{\langle H\rangle_{\rm av}^{1/2}}{\langle H^{1/2}\rangle_{\rm av}^{1/2}} \frac{1}{H(y_0)^{1/2}} \cdot \frac{1}{\hat{\V}_{\blue{E}}^{2/3}} 
    \cdot m_s^{\blue{E_4}}
    = \frac{1}{\langle H\rangle_{\rm av}^{1/2}}\cdot \frac{e^{\langle\Phi\rangle}}{\sqrt{4\pi}H(y_0)^{1/2}\hat{\V}_{\blue{E}}^{2/3}} \cdot\Mp
    }  \,.
\end{equation}

\section{Flux scalar potential}
\label{sec:scalarpotential}

The scalar potential for the moduli fields and the dilaton comes from 
the terms $R$, $|G_3|^2$ and $|\Tilde{F}_5|^2$ in the action (\ref{eq:TypeIIB}), after dimensional reduction to 4d. The contributions from the $R$ and $\Tilde{F}_5$ terms can be shown to give (see section 5.3 of \cite{DeWolfe:2002nn})
\begin{align}\label{eq:RF5contrib}
    \frac{1}{2\kappa^2}\int 
    \left(R\star 1 - \frac{e^{2\Phi_0}}{4}\Tilde{F}_5\wedge\star\Tilde{F}_5\right)
    =\frac{e^{\Phi_0}}{2\kappa^2}\int d^4x\sqrt{-g_4}
    \cdot e^{4\omega(x)}\int H^{-1}\frac{G_3\wedge i\overline{G}_3}{2(\imtau)}\,,
\end{align}
which we can put together with the $G_3\wedge\star\overline{G}_3$ term to give in total
\begin{align}
    S_{\rm IIB}^{\blue{E}} &\supset \frac{e^{\Phi_0}}{2\kappa^2}
    \int d^4x\sqrt{-g_4} \cdot e^{4\omega(x)}
    \int \frac{H^{-1}}{2(\imtau)}G_3\wedge(i\overline{G}_3 + \star_6\overline{G}_3) \\
    &= \frac{e^{\Phi_0}}{2\kappa^2}
    \int d^4x\sqrt{-g_4} \cdot e^{4\omega(x)}
    \int \frac{H^{-1}}{(\imtau)} G_3^+\wedge\star_6\overline{G}_3^+ \,,
    \label{eq:DimRedAppendix_potentialG3term}
\end{align}
with $G_3^+ = \frac{1}{2}(G_3 + i\star_6 G_3)$ such that $\star_6 G_3^+ = -iG_3^+$ 
 \cite{DeWolfe:2002nn}.
Using the metric (\ref{eq:10dmetric}) we can rewrite this action in terms of $g_{mn}$, 
\begin{equation}
    S_{\rm IIB}^{\blue{E}} \supset \int d^{4}x\sqrt{-g_{4}} ~
    \left\{ \frac{e^{\Phi_0}}{2\kappa^2} ~ e^{4\omega(x)}
    \int ~ H^{-1} ~ \frac{G_3^+\wedge\star_{g_6}\overline{G}_3^+}{(\text{Im}\tau)}
    \right\} 
    \equiv -\int d^4x \sqrt{-g_4} ~V \,,
\end{equation}
which defines the 4d scalar potential $V$ as
\begin{equation}
   \boxed{ V = -i~\frac{e^{\Phi_0}}{2\kappa^2} ~ e^{4\omega(x)}
    \int ~ H ~ \frac{(H^{-1}G_3^+) \wedge (H^{-1}\overline{G}_3^+)}{(\text{Im}\tau)}  } \,.
\end{equation}
\noindent It is now possible to rewrite this potential in an ${\cal N}=1$ supergravity form, by defining \cite{Gukov:1999ya} 
\be\label{eq:Wdef}
W= \frac{1}{a} \int G_3\wedge \Omega  \,,
\ee 
where $a$ is a normalisation constant to be determined below, and using that
\be
  \int  ~ H ~ (H^{-1}G_3^+) \wedge (H^{-1}\overline{G}_3^+) = \frac{a^2}{\int~H~\Omega\wedge\overline{\Omega}} G^{\alpha\Bar{\beta}}(D_\alpha W)(D_{\Bar{\beta}}\overline{W}) \,
\ee
where $\alpha,\beta$ run over the complex structure moduli and the axio-dilaton. 
Using this, the scalar potential becomes
\begin{align}
    V &= -i~\frac{e^{\Phi_0}}{2\kappa^2} 
    ~ \frac{e^{4\omega(x)}}{(\Im\tau)}
    \frac{a^2}{\int H~\Omega\wedge\overline{\Omega}} \left( 
    G^{i\Bar{\jmath}}(D_i W)(D_{\Bar{\jmath}}\overline{W}) - 3|W|^2 \right) \nonumber \\
    &= \frac{e^{\Phi_0}}{2\kappa^2} 
    \Big(\frac{e^{2\omega_0}\cdot l_s^6}{V_{\rm w}}\Big)^2\frac{1}{(\Im\tau)}
    \frac{a^2}{l_s^6}
    \frac{l_s^6}{i\int H~\Omega\wedge\overline{\Omega}} \left( 
    G^{i\Bar{\jmath}}(D_i W)(D_{\Bar{\jmath}}\overline{W}) - 3|W|^2 \right) \nonumber \\
    &= \frac{2\pi}{e^{\Phi_0}\cdot l_s^8} 
    \Big(\frac{e^{2\Phi_0}M_\text{Pl}^2 \cdot l_s^2\cdot l_s^6}{4\pi V_{\rm w}}\Big)^2\frac{1}{(\Im\tau)}
    \frac{a^2}{l_s^6}
    \frac{l_s^6}{i\int H~\Omega\wedge\overline{\Omega}} \left( 
    G^{i\Bar{\jmath}}(D_i W)(D_{\Bar{\jmath}}\overline{W}) - 3|W|^2 \right) \nonumber \\
    &= \frac{e^{3\Phi_0}}{4\pi \cdot l_s^4} 
    \Mp^4 
    \frac{a^2}{l_s^6}
    \cdot
    \Big(\frac{l_s^6}{V_{\rm w}}\Big)^2
    \frac{1}{2(\Im\tau)}
    \frac{l_s^6}{i\int H~\Omega\wedge\overline{\Omega}} 
    \cdot\Mp^2\left( 
    \frac{G^{i\Bar{\jmath}}}{\Mp^2}(D_i W)(D_{\Bar{\jmath}}\overline{W}) - \frac{3}{\Mp^2}|W|^2 \right) \nonumber \\
    &= \Bigg\{\frac{e^{3\Phi_0}}{4\pi \cdot l_s^{10}} 
    M_\text{Pl}^6 \Bigg\}  
    e^{K/\Mp^2} 
    \left(K^{i\Bar{\jmath}}(D_i W)(D_{\Bar{\jmath}}\overline{W}) - \frac{3}{\Mp^2}|W|^2 \right) \,,
\end{align}
where now $i,j$ run over complex structure moduli, K\"ahler moduli and the axio-dilaton, the K\"ahler potential $K$ is given by 

\begin{equation}
    K/\Mp^2 = -2\log\V_{\rm w} - \log(-i(\tau-\bar{\tau}))
    - \log\left(\frac{i}{l_s^6}\int H~\Omega\wedge\overline{\Omega}\right)
\end{equation}
and $K^{i\Bar{\jmath}}$ is the inverse field space metric that follows from $K_{i\Bar{\jmath}} = \partial_i\partial_{\Bar{\jmath}} K$. 

Note that the volume term in $K$ includes not only the overall volume modulus $\V$, but also the other K\"ahler moduli. This scalar potential leads to the normalisation
\begin{equation}
    W/\Mp^3 = \frac{e^{\frac{3}{2}\Phi_0}}{\sqrt{4\pi} \cdot l_s^5} \int G_3\wedge \Omega \,.
\end{equation}
We can see that the normalisation constant, $a$, is convention-dependent through the choice of $e^{\Phi_0}$. 

This gives for the gravitino mass 
\begin{equation}\label{eq:mgravitino}
    \boxed{m_{3/2} = e^{\frac{K}{2\Mp^2}} \frac{|W|}{\Mp^2} = \frac{e^{\frac{1}{2}\langle\Phi\rangle}}{\V_{\rm w}~||\Omega||_{\rm w}} \frac{e^{\frac{3}{2}\Phi_0}W_0}{\sqrt{8\pi}} \Mp} \,, 
\end{equation}
where $e^{\frac{1}{2}\langle\Phi\rangle}$ comes from $\langle\Im\tau\rangle$, $||\Omega||^2_{\rm w}\cdot l_s^6 = i\int H~\Omega\wedge\overline{\Omega}$ and we define 
\begin{equation}
    W_0/\Mp^3 \equiv \left\langle \frac{1}{l_s^5}\int G_3\wedge\Omega \right\rangle \,.
\end{equation}
It follows from \eqref{eq:mgravitino} and \eqref{eq:mKK} that the important\footnote{Not only is this ratio important because a consistent 4d supergravity description requires that the gravitino remains in the theory, i.e. its mass is not above the EFT cutoff -- typically $m_{KK}$ -- and therefore integrated out, but it was shown that it also serves as a control parameter for certain corrections to the scalar potential, e.g. from higher F-terms \cite{Cicoli:2013swa}.} ratio (assuming the bulk dominates all the integrals, so that $\V_{\rm w}\approx\V$ and $||\Omega||_{\rm w}\approx ||\Omega||$) 
\begin{equation}
    \frac{m_{3/2}}{m_{KK}} = 
    H_0^{1/2}~\frac{e^{\frac{1}{2}(\langle\Phi\rangle+\Phi_0)}}{\V_{\blue{E}}^{1/3}} 
    \frac{W_0}{\sqrt{2}(2\pi)||\Omega||} \,,
\end{equation}
where we highlight the fact that the volume being used is the \blue{Einstein frame} volume in \red{string frame} string length units, $\V_{\blue{E}}$. 
A manifestly convention-independent relation is found by expressing the \blue{Einstein frame} volume in \blue{Einstein frame} string length units, which gives
\begin{equation}
    \frac{m_{3/2}}{m_{KK}} = 
    H_0^{1/2}~\frac{e^{\langle\Phi\rangle}}{\hat{\V}_{\blue{E}}^{1/3}} 
    \frac{W_0}{\sqrt{2}(2\pi)||\Omega||} \,.
\end{equation}

\section{Corrections to the scalar potential} 
\label{sec:corrections}

Since the flux superpotential leaves all K\"ahler moduli unstabilised, either leaving them as flat directions or generating runaways, one must resort to higher-order corrections to the EFT in order to stabilise them. Both perturbative and non-perturbative corrections have been considered in the literature --- while the former are computed at the level of the 10d EFT and in \red{string frame}, the latter are obtained directly at the level of the 4d EFT and are computed in \blue{Einstein frame}. One must therefore be careful with the conventions being used to change frames, i.e. the choice of $\Phi_0$, in order to remain consistent. 

\subsection{Perturbative corrections}

In \cite{Becker:2002nn}, it was shown that $\alpha'$-corrections to the Type IIB effective action (\ref{eq:TypeIIB}) manifest as corrections to the 4d volume modulus K\"ahler potential and spoil the no-scale structure of its scalar potential. These corrections arise from higher-derivative terms at order $(\alpha')^3$ appearing in the type IIB effective action,
\begin{equation}
    S^{\red{S}}_{\rm IIB} = \frac{1}{2\kappa_{10}^2}\int d^{10}x \sqrt{-G^{\red{S}}} ~e^{-2\Phi}\Big(R^{\red{S}} + 4(\partial\Phi)^2_{\red{S}} + (\alpha')^3\cdot\frac{\zeta(3)}{3\cdot 2^{11}}\cdot J_0\Big) \,,
\end{equation}
where the higher-order term is schematically given by
\begin{equation}
    J_0 \sim (R_{MNPQ})^4 \,.
\end{equation}
One must also add a term 
\begin{equation}
    \delta S^{\red{S}}_{\Phi} \sim \int d^{10}x \sqrt{-G^{\red{S}}} e^{-2\Phi}(\alpha')^3 (\nabla^2\Phi)~Q \,,
\end{equation}
where $Q\sim (R_{MNPQ})^3$ is a generalisation of the 6d Euler integrand $\int_{X_{6}} d^6y\sqrt{g_6}~Q = \chi$, with $\chi$ the Euler characteristic of $X_6$ \cite{Becker:2002nn}.
This term corrects the 10d solution to the equation of motion for $\Phi$, such that $\Phi = \Phi_{10} + \frac{\zeta(3)}{16}Q$. It is then shown in \cite{Becker:2002nn} that this leads to a correction to the K\"ahler potential of the form
\begin{align}
    K &= -2\log\Big(\V_{\red{S}} + \frac{\xi}{2}\Big)
    = -2\log\Big(\V_{\blue{E}}~e^{\frac{3}{2}(\Phi - \Phi_0)} + \frac{\xi}{2}\Big) \\
    &= -2\log\Big(\V_{\blue{E}} + \frac{\xi}{2}e^{-\frac{3}{2}(\Phi-\Phi_0)}\Big) 
    + ... \,,
\end{align}
where we have used (\ref{eq:Vd_between_frames}) with $d=6$, keeping $\Phi_0$ unspecified, and $\xi$ is defined as\footnote{In \cite{Becker:2002nn}, we find the definition $\xi = -\frac{\zeta(3)\chi(X_6)}{2}$. The missing factor of $(2\pi)^3$ comes from their conventions for the volume, $\V_{\text{\tiny\cite{Becker:2002nn}}} = V_6/(2\pi\alpha')^3$, whereas we are using $\V = V_6/l_s^6 = (2\pi)^{-3}~V_6/(2\pi\alpha')^3$, with the convention $(2\pi)^2\alpha' = l_s^2$. There are also instances in the literature where the factor of $1/2$ is absorbed into the definition of $\xi$.}
\begin{equation}
    \xi = -\frac{\zeta(3)\chi}{2(2\pi)^3} \,.
\end{equation}
Note in particular that the correction expressed in \blue{Einstein frame} depends on the convention one chooses for $\Phi_0$,
\begin{align}
    \Phi_0=0 &\implies K = -2\log\left(\V + \frac{\xi}{2g_s^{3/2}}\right) \,, \\
    \Phi_0 = \vev{\Phi} &\implies K = -2\log\left(\V + \frac{\xi}{2}\right) \,,
\end{align}
where we have assumed as usual that the dilaton has been stabilised by fluxes.

\subsection{Non-perturbative corrections}

Although the superpotential $W$ does not receive perturbative corrections, it may receive non-perturbative corrections from either instantons arising from Euclidean D3-branes wrapping 4-cycles or gaugino condensation on the world-volume theory of D7-branes wrapped around internal 4-cycles. 
Let us consider the latter case in some detail. In what follows, $T_p = \frac{2\pi}{l_s^{p+1}}$ is the brane tension. The DBI action  for a Dp-brane, in the \red{string frame}, is given by \cite{Polchinski:1998rr,Becker:2007zj}
\begin{align}
    \action{Dp}{DBI} = - T_p \int d^{p+1}\sigma ~e^{-\Phi} \sqrt{-\det(g^{\red{S}} + B + \frac{l_s^2}{2\pi}F)} \,,
\end{align}
where $g^{\red{S}}$ and $B$ refer to the pull-back of the \red{string frame} metric $(G^{\red{S}})_{MN}$ and 2-form $B_{MN}$ onto the world-volume of the brane and $F$ to the field-strength $F_{ab}$ of the brane gauge fields. Rewriting the action in terms of the \blue{Einstein frame} metric,
\begin{align}
    \action{Dp}{DBI} &= - T_p \int d^{p+1}\sigma ~e^{-\Phi} 
    \sqrt{-\det g^{\red{S}}}
    \sqrt{\det(\mathbf{1} + (g^{\red{S}})^{-1}\Big(B + \frac{l_s^2}{2\pi} F\Big))}  \\
    &\supset - T_p \int d^{p+1}\sigma ~e^{-\Phi} 
    \sqrt{-\det g^{\red{S}}} 
    ~\frac{1}{4}\Big(\frac{l_s^2}{2\pi}\Big)^2(g^{\red{S}})^{ac}(g^{\red{S}})^{bd} F_{ab}F_{cd}  \\
    &=  - \frac{T_p}{4} \frac{l_s^4}{(2\pi)^2}\int d^{p+1}\sigma ~
    \sqrt{-\det g^{\blue{E}}} ~e^{\frac{p-3}{4}(\Phi-\Phi_0)}e^{-\Phi} F_{ab}F^{ab} \,,
\end{align}
where the indices in $F_{ab}F^{ab}$ are contracted with \blue{Einstein frame} metrics. This is the kinetic term for the brane gauge bosons (see Appendix A.2. of \cite{Parameswaran:2020ukp}) and tells us the gauge coupling of the corresponding theory, which is a key parameter for gaugino condensation. If the brane is wrapping a $(p-3)$--cycle $\Sigma_{p-3}$, we find the corresponding 4d term (assuming that $\Phi$ is constant over the cycle)
\begin{align}
    \action{Dp}{4d} &\supset - \frac{1}{8\pi l_s^{p-3}}\int d^{4}x ~
    \sqrt{-\det g^{\blue{E}}_4} ~e^{\frac{p-3}{4}(\Phi-\Phi_0)}e^{-\Phi} \underbrace{\left(\int d^{p-3}\sigma \sqrt{g^{\blue{E}}_{p-3}} \right)}_{\tau_{\Sigma_{p-3}}^{\blue{E}} l_s^{p-3}} F_{ab}F^{ab} \\
    &= - \int d^{4}x ~\sqrt{-\det g^{\blue{E}}_4}
    \left\{\frac{\tau_{\Sigma_{p-3}}^{\blue{E}}}{8\pi e^{\Phi}} ~e^{\frac{p-3}{4}(\Phi-\Phi_0)}\right\} F_{ab}F^{ab} \,,
\end{align}

\noindent and we can read off the gauge coupling $g_c$, 

\begin{equation}
    \boxed{\frac{1}{g_c^2} = \frac{\tau_{\Sigma_{p-3}}^{\blue{E}}}{4\pi e^{\vev{\Phi}}} ~e^{\frac{p-3}{4}(\vev{\Phi}-\Phi_0)}} \,,
\end{equation}

\noindent where we have assumed as usual that the dilaton has been stabilised by fluxes at some higher scale.  Gaugino condensation on the world-volume theory of D7-branes will then give a non-perturbative contribution to the superpotential \cite{Hebecker:2020aqr},\footnote{Here $N$ is the number of branes stacked on top of each other, responsible for the gauge group. It appears through the beta-function coefficient \cite{Hebecker:2020aqr}.} 
\begin{align}
    W_{np} \sim e^{-\frac{8\pi^2}{g_c^2}\frac{1}{N}} = e^{-\frac{2\pi}{N}\frac{\tau^{\blue{E}}}{g_s}e^{\vev{\Phi}-\Phi_0}} \,,
\end{align}
where we used $e^{\vev{\Phi}} = g_s$. Holomorphicity of $W$ then leads to the general contribution
\begin{equation} \label{eq:Wnp}
    \boxed{W_{np} = \sum_i^{\phantom{i}} A_i e^{i\frac{a_i}{g_s} e^{\vev{\Phi}-\Phi_0} T_i^{\blue{E}}}} \,,
\end{equation}
where the sum runs over the contributing cycles, $a_i = \frac{2\pi}{N_i}$ and the fields $T_i^{\blue{E}} = b_i^{\blue{E}} + i\tau_i^{\blue{E}}$,  are the complexified K\"ahler moduli. Hence, we can compare the two most common conventions used for $\Phi_0$,
\begin{align}
    \Phi_0=0 &\implies W_{np} = \sum_i A_i e^{ia_i T_i^{\blue{E}}} \,, \\
    \Phi_0 = \vev{\Phi} &\implies W_{np} = \sum_i A_i e^{i\frac{a_i}{g_s} T_i^{\blue{E}}} \,.
\end{align}
As for the mass ratio $m_{3/2}/m_{KK}$, the superpotential as written, in terms of \blue{Einstein frame} 4-cycle volumes in \red{string frame} string length units, appears to depend on the choice of convention for $\Phi_0$, but one should recall that the 4-cycle volumes also depend on this choice of convention.  If we express the 4-cycle volumes in terms of the \blue{Einstein frame} string length units (\ref{eq:Vd_between_frames}), 
\begin{equation}
    \hat{\tau}_i^{\blue{E}} = e^{\vev{\Phi}-\Phi_0} \tau_i^{\blue{E}} \,,
\end{equation}
the convention-independence of \eqref{eq:Wnp} becomes manifest.

\section*{Acknowledgements}
We are grateful to Anshuman Maharana for helpful discussions.
IZ is partially supported by STFC, grant ST/P00055X/1.  SLP is partially supported by STFC, grant ST/X000699/1.

\smallskip
\noindent For the purpose of open access, the authors have applied a Creative Commons Attribution (CC BY) licence to any Author Accepted Manuscript version arising. Data access statement: no new data were generated for this work.
\bibliographystyle{JHEP.bst}
\bibliography{ref}

\end{document}